\documentclass[aps,fleqn,pre,showpacs,twocolumn]{revtex4}
\usepackage{amsmath,amssymb,graphicx}

\newcommand{\mr}{\mathrm}
\newcommand{\mt}{\text}

\newcommand{\mean}[1]{\langle{#1}\rangle}

\newcommand{\km}{k_{\mathrm{min}}}

\newcommand{\kM}{k_{\mathrm{max}}}
\newcommand{\knn}{k_{\mathrm{nn}}}
\newcommand{\knni}[1][i]{k^{(i)}_{\mathrm{nn}}}
\newcommand{\knnl}[1][i]{k^{(l)}_{\mathrm{nn}}}

\newcommand{\tknni}[1][i]{\tilde{k}^{(i)}_{\mathrm{nn}}}

\newcommand{\p}[2]{P^{\mr{#2}}_{\mr{#1}}}
\newcommand{\pe}{\p{e}{}}
\newcommand{\pkd}{\p{}{(d)}}
\newcommand{\ped}{\p{e}{(d)}}

\newcommand{\pjkUc}{P_{\mr{UC}}(j,k)}
\newcommand{\pjkcd}{P^{\mr{(d)}}(j|k)}

\begin{document}

\title{Generation of arbitrarily two-point correlated random networks}

\author{Sebastian~Weber}
\affiliation{Institut~f\"ur~Festk\"orperphysik,
             Technische~Universit\"at~Darmstadt,
             Hochschulstr.~8, 64289~Darmstadt, Germany}

\author{Markus~Porto}
\affiliation{Institut~f\"ur~Festk\"orperphysik,
             Technische~Universit\"at~Darmstadt,
             Hochschulstr.~8, 64289~Darmstadt, Germany}

\date{\today}

\begin{abstract}
  Random networks are intensively used as null models to investigate
  properties of complex networks. We describe an efficient and
  accurate algorithm to generate arbitrarily two-point correlated
  undirected random networks without self- or multiple-edges among
  vertices. With the goal to systematically investigate the influence
  of two-point correlations, we furthermore develop a formalism to
  construct a joint degree distribution $P(j,k)$ which allows to fix
  an arbitrary degree distribution $P(k)$ and an arbitrary average
  nearest neighbor function $\knn(k)$ simultaneously.  Using the
  presented algorithm, this formalism is demonstrated with scale-free
  networks ($P(k) \propto k^{-\gamma}$) and empirical complex networks
  ($P(k)$ taken from network) as examples. Finally, we generalize our
  algorithm to annealed networks which allows networks to be
  represented in a mean-field like manner.
\end{abstract}


\maketitle

\section{Introduction}
\label{sec:introduction}

The fast developing research field of complex networks
\cite{Albert2002,Dorogovtsev2002} focuses on the three main aspects of
(i) measuring network topology, (ii) investigating dynamics on
networks, and (iii) studying the interplay between dynamical processes
on networks and the network topology. Surprisingly, empirical networks
from a vast variety of scientific fields share a lot of
characteristical features.  Prominent examples are the small-world
property \cite{Watts1998}, high clustering \cite{Newman2003b}, and the
scale-free degree distribution \cite{Barabasi1999}. One possibility to
unravel the properties of empirical networks is to compare them to
null models. Appropriate null models are random networks with some of
the statistical features preserved being present in the empirical
network under investigation. This idea gave birth to the well-known
configuration model (CM) algorithm
\cite{Bender1978,Bollobas1980,Molloy1995,Molloy1998,Catanzaro2005}
which is capable of generating random networks with an {\em a priori}
given degree distribution. Some extensions to this model have been
proposed to even conserve some further statistical properties than the
plain degree distribution, for instance the degree dependent
clustering coefficient \cite{Serrano2005}.

A fundamental way to categorize and distinguish empirical networks
beyond the degree distribution and clustering has been proposed by
\citeauthor{Newman2002} \cite{Newman2003,Newman2002} who introduced
the Newman factor $r$.  This number is basically the Pearson
correlation coefficient of degrees (the number of edges emanating from
a vertex) from connected vertices in a network and is therefore fully
defined by two-point correlations in a network. The range of the
Newman factor is in the interval $[-1,1]$ where positive (negative)
values indicate that vertices with the same (different) degree tend to
be connected, while a value of $0$ means no correlation. Practically
all empirical networks show a non-trivial two-point correlation
structure. An astonishing observation is, for example, the fact that
biological networks show negative Newman factors, while
technological networks display rather small values of the Newman
factor close to zero, whereas social networks tend to have
rather large positive values \cite{Newman2003a}. The evident
importance of correlations within the degree distribution has led to
lots of efforts, for example a hidden variable approach has been
developed in Ref.~\cite{Bogun'a2003a} and so-called $dK$-series
networks which systematically describe the full correlation structure
of a network have been introduced in Ref.~\cite{Mahadevan2006}
together with an algorithm for the lowest $dK$-classes. Thus, an
efficient random network generator which constructs null model
networks at the basis of an {\em a priori} prescribed two-point
correlation structure is very important. Such a generator is presented
below and allows to construct undirected random networks with a
prescribed two-point correlation structure and hence much more
realistic null models.  The major advantage of our generator in
  comparison with similar algorithms previously introduced
  \cite{Bogun'a2003a,Mahadevan2006,V'azquez2003} is its high accuracy
  and the generality of the approach which allows to construct
  networks with an arbitrary two-point correlation structure.  As an
  application of this scheme and in order to investigate the
influence of two-point correlations within empirical networks, we
address the question how one can model two-point correlations while
preserving the degree distribution of a network.  This is fundamental,
for instance, in order to shed light on the interplay between
dynamical processes on networks on the underlying network topology
with respect to two-point correlations.

The modeling of two-point correlations is especially interesting for
the verification of theoretical predictions from theories describing
dynamical processes on networks which do incorporate two-point
correlations. Due to the small-world effect present in networks, it is
common use to utilizes a mean-field (MF) ansatz. Hence, within these
theories the network is modeled using a probabilistic approach and
vertices are only connected with a certain probability to each other.
The idea to represent a network by probabilities has already been
brought up in the context of Kauffman's model of random complex
automata \cite{Derrida1986,Bastolla1996}.  This so-called annealed
network changes in every time step such that all edges are
redistributed. A similar approach has recently been applied by
\citeauthor{Stauffer2005} to scale-free networks to study the effect
of `annealed disorder' on a diffusion process~\cite{Stauffer2005}.
Such annealed networks are ideally suited to test the validity of MF
theories of dynamics on networks. We extend this approach below by
generalizing our algorithm to allow for the construction of two-point
correlated annealed networks.

This paper is organized as follows: Section II introduces the network
correlation measures used in this paper. Section III describes the
algorithm to construct arbitrarily two-point correlated networks.
Section IV develops a formalism which allows to fix a degree
distribution and to arbitrarily choose the two-point correlations at
the same time. The formalism is demonstrated with scale-free networks
and empirical networks as examples. Section VI introduces the notion
of a two-point correlated annealed network. We conclude and give an
outlook in section VII.

\section{Correlation Measures}
\label{sec:corNetworks}

The following is a short summary of common definitions adapted to our
purposes which will be used frequently within this paper.  Two-point
correlations are statistically described by the joint degree
distribution $P(j,k)$ which is the probability that a randomly chosen
edge of the network has vertices with degrees $j$ and $k$ at its ends.
This distribution is a symmetric function in the case of undirected
networks, $P(j,k) = P(k,j)$. By summation over either parameters of
$P(j,k)$, one obtains the distribution over edge ends,
\begin{equation}
  \label{eq:pkeDef}
  \pe(k) = \sum_{j} P(j,k) ,
\end{equation}
which is related to the distribution of vertices by
\begin{equation}
  \label{eq:pkVertDef}
  P(k) = \frac{\bar{k}}{k} \pe(k).
\end{equation}
This last relation \eqref{eq:pkVertDef} between the edge end
distribution $\pe(k)$ and the degree distribution $P(k)$ can easily be
understood by the fact that every vertex with degree $k$ has
probability $P(k)$ of being drawn at random from the network.
Therefore, the probability to draw an edge end connected to a vertex
of degree $k$ is proportional to $k P(k)$. Normalizing this last
expression yields the edge end distribution $\pe(k) = k P(k)/\bar{k}$.
Here, $\bar{k} = \sum_k k P(k)$ denotes the mean with respect to the
degree distribution $P(k)$.  This mean has to be carefully
distinguished from the mean with respect to the edge end distribution
$\pe(k)$ which we denote by $\mean{k} = \sum_k k \, \pe(k) =
\overline{k^2}/\bar{k}$.  It is convenient \cite{Song2006} to extract
the actual correlations from $P(j,k)$ by relating it to the
uncorrelated case $\pjkUc$, which has the special product form
\begin{equation}
  \label{eq:pkkUC}
  \pjkUc = \pe(j) \, \pe(k) .
\end{equation}
By taking the ratio between $P(j,k)$ and $\pjkUc $, this defines
\begin{equation}
  \label{eq:fkkDef}
  f(j,k) = \frac{P(j,k)}{\pjkUc}
\end{equation}
as a correlation function.

However, the joint degree distribution $P(j,k)$ and the correlation
function $f(j,k)$ are complex functional objects which are hard to
imagine. A way to quantify the overall correlation present in a
network was introduced by Newman \cite{Newman2002}. He defined the
Newman factor $r$ to be the Pearson correlation coefficient of the
remaining degrees of two vertices at either ends of a randomly chosen
edge. The use of the remaining degree, which is the actual degree of a
vertex minus one, is only an arithmetic trick to suppress some terms
in calculations performed by Newman. In this paper, we directly use
the degrees of the vertices, which is equivalent to Newman's
definition in the limit of large networks,
\begin{equation}
  \label{eq:newmanDef}
  r = \frac{1}{\sigma^2_\mr{e}} \sum_{j,k}  j k ( P(j,k) -
  \pe(j) \pe(k) ) .
\end{equation}
The Newman factor $r$ is normalized by $\sigma^2_\mr{e} =
\mean{k^2} - \mean{k}^2$ to fall into the range $[-1,1]$. A positive
(negative) value means that vertices with a degree $k$ preferentially
attach to vertices with a degree of the same (different) order which
is referred to as (dis-)assortative mixing.  The special case of $r =
0$ is achieved in the case of no correlation, which can be seen by
substituting $\pjkUc$ of Eq.~\eqref{eq:pkkUC} into
Eq.~\eqref{eq:newmanDef}. It is clear that the Newman factor $r$
quantifies the correlations present in a network only on a global
scale.  An intermediate approach, being on the level of degrees, has
been introduced in Ref.~\cite{Bogun'a2003} with the average nearest
neighbor function $\knn(k)$.  Using the conditional probability
\begin{equation}
  \label{eq:condProbDef}
  P(j|k) = \frac{P(j,k)}{\pe(k)} ,
\end{equation}
which is the probability that a randomly chosen neighbor of any vertex
with degree $k$ has the degree $j$, one defines $\knn(k)$ to be
\begin{equation}
  \label{eq:knnDef}
    \knn(k) = \sum_{j} j \, P(j|k) .
\end{equation}
In the case of an (dis-)assortative network the average nearest
neighbor $\knn(k)$ has to be an (de-)increasing function, while it
has the constant value $\mean{k}$ for uncorrelated networks. It is
interesting to note that
\begin{equation}
  \label{eq:knnMeanProp}
  \mean{\knn(k)} = \mean{k}
\end{equation}
is generally valid, which can be seen by plugging
Eq.~(\ref{eq:condProbDef}) into Eq.~(\ref{eq:knnDef}) and averaging
the resulting equality over $k$ with respect to the edge end
distribution $\pe(k)$.

\section{Algorithm}
\label{sec:algorithm}

%
%

The well-known CM algorithm
\cite{Bollobas1980,Bender1978,Molloy1998,Molloy1995} fixes {\em a
  priori} a degree sequence which is usually drawn from a given degree
distribution $P(k)$. Each element of this degree sequence is the
number of desired edges emanating of a vertex. These may be thought of
as half-edges which still need to be joined with half-edges of other
vertices. To construct the network, the CM algorithm may be
implemented by placing all half-edges of all vertices into a single
list, which is a discrete representation of the edge end distribution
$\pe(k)$. An edge is formed by selecting two random members of that
list. If the constraint of neither self- nor multiple-edges is met,
the edge is created and the two half-edges are removed from the list.
As the first and the second draw is done from the same list or,
equivalently, each draw is done independently with the edge end
distribution $\pe(k)$, the resulting network is always uncorrelated.
Only the constraint of self- and multiple-edge prevention induces some
intrinsic correlations, which can be avoided if the maximal degree
$\kM$ is limited (cf.  section~\ref{sec:sfNets}). The CM algorithm
paired with the correct choice of the maximal degree $\kM$ is as well
known as the uncorrelated CM (UCM) algorithm \cite{Catanzaro2005}.
However, almost all empirical networks do display two-point
correlations in their topology. The algorithm discussed below allows
to fix {\em a priori} an arbitrary joint degree distribution
$P(j,k)$ and generates a network which is completely random under all
other topological aspects, just as the CM algorithm does with respect
to the degree distribution $P(k)$.

A major computational complication arises from the fact that
probabilities in the $P(j,k)$ matrix may become very small as the
probability for one edge is of the order $1/\bar{k} N$ and
computationally hard to handle for large $N$. Due to this problem, we
sample in a first step a half-edge with the usual edge end
distribution $\pe(k)$, in a second step, we sample a half-edge from
the conditional probability distribution $P(j|k)$. The former two
objects are much easier to sample as those are the result of integrals
over $P(j,k)$ and therefore contain probabilities of greater order.

The overall scheme of the algorithm to construct a network with $N$
vertices and a given joint degree distribution $P(j,k)$ is the
following:

\begin{enumerate}
\item As in the CM algorithm, one first has to draw a degree sequence
  by calculating the theoretical (continuous) edge end distribution
  $\pe(k)$ from the joint degree distribution $P(j,k)$ and transform
  that into a degree distribution $P(k)$. From this distribution, a
  degree sequence of length $N$ is drawn.
\item Each element of the degree sequence represents a vertex. All
  vertices with the same degree $k$ are then sorted into degree
  classes, each containing only vertices of the same degree $k$.
\item To compensate for discretization effects caused by the
    finiteness of the sampled network, one has to calculate the
  discrete edge end distribution $\ped(k)$ from the generated degree
  sequence. To do so, one acquires, by estimating the size of each
  degree class, the discrete degree distribution $\pkd(k)$, which
  corresponds to a discrete edge end distribution by $\ped(k) = k \,
  \pkd(k) /\bar{k}$.
\item Next, the discrete conditional probability $\pjkcd$ is setup. To
  obtain a matrix which accommodates the discretization effects, one
  replaces the continuous edge end distributions $\pe(k)$ in the
  definition of the conditional probability distribution of
  Eq.~(\ref{eq:condProbDef}) by the discrete edge end distributions
  $\ped(k)$ and obtains therefore
  \begin{equation}
    \label{eq:2}
    \begin{split}
      P(j|k) &= \frac{P(j,k)}{\pe(k)} = \pe(j) \, f(j,k) \\
      &\approx \ped(j) \, f(j,k) =
      \ped(j) \, \frac{P(j,k)}{\pe(j) \, \pe(k)} .
    \end{split}
  \end{equation}
  Since we mix the discrete edge end distribution $\ped(j)$ and the
  continous correlation function $f(j,k)$, the resulting conditional
  degree distribution $\pjkcd$ is only approximately normalized for a
  given degree class $k$. To obtain a conditional probability
  distribution suitable for sampling degree classes, we normalize each
  degree class separately, leading to the final form
  \begin{equation}
    \label{eq:3}
      \pjkcd = \frac{\ped(j)}{\pe(j)} \, P(j,k) \,
      \left(\sum_j \frac{\ped(j)}{\pe(j)} \, P(j,k) \right)^{-1} .
  \end{equation}
  This definition is consistent with the limes $N~\rightarrow~\infty$,
  as the discrete edge end distribution $\ped(j)$ becomes equal in
  this limit to the continous edge end distribution $\pe(j)$ and the
  ratios $\ped(j)/\pe(j)$ become exactly $1$, respectively.
\item After all base data structures have been initialized, the
  algorithm starts to draw edges by drawing edge ends. The first edge
  end is selected by first drawing a degree class $k$ from the edge
  end distribution $\ped(k)$ and then randomly choose a vertex from
  that degree class.
\item The second end of the edge is chosen in the same two step
  manner.  However, the first draw of a degree class $j$ is done with
  the appropriate conditional probability distribution $\pjkcd$
  instead of the edge end distribution $\ped(k)$. This construction
  scheme yields correctly correlated graphs, since we have
  \begin{equation}
    \label{eq:1}
    \underbrace{\pe(k)}_{\text{1. draw}} \,
    \underbrace{P(j|k)}_{\text{2. draw}} = P(j,k) .
  \end{equation}
  An edge is created whenever the constraints of neither self- nor
  multiple-edges is met. Otherwise the drawn edge is rejected and
  the algorithm continues with step five.
\item If the edge is created, the probability weights of the two edge
  ends are removed from the corresponding degree classes in the edge
  end distribution $\ped(k)$ and the conditional probability
  distribution matrix $\pjkcd$. The removal of the probability weight
  is equivalent to the removal of the two half-edges from the list of
  eligible half-edges in the CM algorithm.
\item The steps five to seven are repeated until no edge ends are left
  and all edges are formed.
\end{enumerate}

The principal numerical costs of the algorithm arises from the
continuous sampling of degree classes in the steps five and six above.
Since the algorithm has to sample only the degree classes actually
realized, which is a significant lower number than the system size
$N$, the numerical costs are of the order $\mathcal{O}(N^\alpha)$ with
$\alpha < 1$.  Furthermore, due to the removal of probability weight
of used half-edges throughout the construction procedure, the
algorithm samples only the possible configuration space which remains
valid in each iteration step just as in the CM algorithm. The memory
usage of the algorithm scales with the square of the number of
realized degree classes. This can become a significant advantage over
the CM procedure as described above, since the memory usage of the CM
procedure scales with the number of half-edges needed to construct the
network.

\begin{figure}[t]
  \centering
  \includegraphics{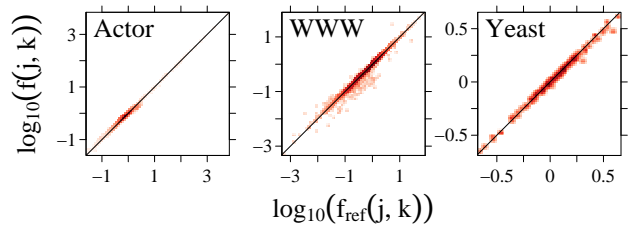}
  \caption{(color online) Density plot of the correlation function
    $f_{\text{ref}}(j,k)$ of the empirical network versus the
    correlation function $f(j,k)$ of the corresponding random network
    as generated by the algorithm for all indices $j$ and $k$.
    Darker red regions contain a higher density of data points, while
    lighter red indicates a lower density. The reference line $y = x$
    is drawn as a guide to the eye.}
  \label{fig:corCoeff}
\end{figure}

\begin{figure}[t]
  \centering
  \includegraphics{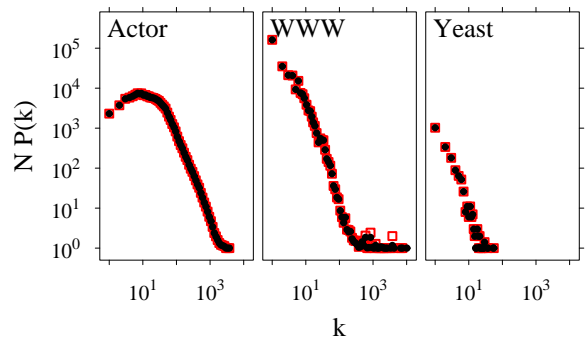}  
  \caption{(color online) Degree distribution $P(k)$ of empirical
    networks and their corresponding degree distribution as generated
    by the algorithm. The red squares denote the reference points as
    measured from the empirical networks and the black circles mark
    values measured from the randomized networks.}
  \label{fig:pkCorr}
\end{figure}

To validate our algorithm, we use three empirical networks as test
cases: (i) a social network where the $392,340$ vertices are actors
and the edges between those are assigned if they performed in at least
one movie together \cite{Barabasi1999}; (ii) a subset of the WWW
containing $325,759$ web pages which are connected if there exists a
link among them \cite{Albert1999}; (iii) the yeast protein-interaction
network constituent of $1,846$ proteins \cite{Jeong2001}. The data has
been downloaded from Barab{\'a}si's web site
\mbox{\url{http://www.nd.edu/~networks}}. All self- and multiple-edges
were removed from each network. The actor network is assortatively ($r
= 0.27$), the WWW network weakly ($r = -0.053$) and the yeast
protein-interaction network disassoartively ($r = -0.16$) correlated.
To test the correctness of the algorithm, one measures the joint
degree distribution $P_{\text{ref}}(j,k)$ of the base networks and
uses this function as input for the construction algorithm. The
resulting random network has to display the same degree distribution
$P(k)$ and joint degree distribution $P(j,k)$ as the empirical one. A
very sensitive test to validate if the correlation structure of the
reference and the random network indeed match is on the level of the
correlation function $f(j,k)$ which varies on a much smaller scale
than the joint degree distribution $P(j,k)$. Thus, comparing the
reference correlation function $f_{\text{ref}}(j,k)$, which one
obtains from the empirical network, with the correlation function
$f(j,k)$ of the network as generated by the algorithm by means of a
correlation coefficient ($1$ means total agreement, $-1$ indicates
that the two functions are of opposite sign and $0$ means no
correlation among the two functions in comparison) reveals almost
complete agreement of (i) $0.99(6)$ (ii), $0.9(9)$, and (iii)
$0.99(8)$. A density plot of the reference correlation function versus
the resulting correlation function in Fig.~\ref{fig:corCoeff} verifies
the excellent agreement of the correlation functions $f(j,k)$ and
$f_{\text{ref}}(j,k)$. The plot shows the corresponding values of
$f(j,k)$ versus $f_{\text{ref}}(j,k)$ for all indices $j$ and $k$ at
either axis.  Ideally, all data-points would be on the diagonal which
would be the case if the two functions were identical and the density
plot would show a delta-shaped line along the diagonal. As one can see
from the plots, the highest density of points, which is indicated by
darker red, is almost solely centered at the diagonal. Just as the
correlation functions coincide, the degree distributions show the same
very good agreement, which is illustrated in Fig.~\ref{fig:pkCorr}.
The statistics per curve are $10^2$ randomized realizations for the
actor-, $10^3$ for the WWW- and $10^4$ for the yeast-network in both
figures.

%
%

\section{Controlling Correlations in Networks}
\label{sec:corrNets}

The algorithm described in this paper constructs undirected random
networks with an arbitrary two-point correlation structure. This
allows us to test explicitly the influence of two-point correlations
present in a network on its properties. For example, being able to
control the two-point correlation structure of a network allows to
directly test their influence on dynamical processes taking place on
the networks. We therefore aim at developing a formalism which allows
to control the two-point correlations of a whole network in terms of
the average nearest neighbor degree $\knn(k)$ and the Newman factor
$r$, given a fixed degree distribution $P(k)$.

As we want to preserve a given degree distribution $P(k)$, which
translates into a given edge end distribution $\pe(k)$, while varying
the joint degree distribution $P(j,k)$, some restrictions apply to the
joint degree distribution. We begin with an ansatz by writing the
joint degree distribution $P(j,k)$ in product form as in
Eq.~(\ref{eq:fkkDef}),
\begin{equation}
  \label{eq:pjkAnsatz}
  P(j,k) = \pe(j) \, \pe(k) \, f(j,k).
\end{equation}
It is clear that the correlations in the network are encoded by this
ansatz within the correlation function $f(j,k)$. The relation to the
Newman factor $r$ from the definition Eq. \eqref{eq:newmanDef} is
\begin{equation}
  \label{eq:newmanFjkRel}
  r \sigma^2_\mr{e} = \mean{jk \, (f(j,k) - 1)}_{j,k} = \mean{jk \, f(j,k)}_{j,k} - \mean{k}^2.
\end{equation}
By the notation $\mean{\cdot}_{j,k}$, we indicate that the average
with respect to $\pe(k)$ is to be taken simultaneously over the
indices $j$ and $k$, similarly as $\mean{\cdot}$ denotes the average
with respect to $\pe(k)$.  The correlation function $f(j,k)$ is as
well tightly connected to the average nearest neighbor degree function
$\knn(k)$. Using that the conditional probability $P(j|k) =
P(j,k)/\pe(k) = \pe(j) \, f(j,k)$, the definition of Eq.
\eqref{eq:knnDef} turns into
\begin{equation}
  \label{eq:knnFjkRel}
  \knn(k) = \mean{j \, f(j,k)}_j .
\end{equation}
Multiplying the average nearest neighbor function $\knn(k)$ with $k\,
\pe(k)$ and summing over all $k$, we are lead to
\begin{equation}
  \label{eq:knnLemma}
  \mean{k \, \knn(k)} = \mean{jk \, f(j,k)}_{j,k} ,
\end{equation}
which we can substitute into Eq. \eqref{eq:newmanFjkRel}, leading us
finally to
\begin{equation}
  \label{eq:knnNewmanRel}
  r \sigma^2_\mr{e} = \mean{k \, \knn(k)} - \mean{k}^2 .
\end{equation}
From the constraint of a given degree distribution $P(k)$ it follows
that an integration over either argument of the joint degree
distribution $P(j,k)$ has to be equal to the corresponding edge end
distribution $\pe(j)$ (or $\pe(k)$). Thus, the correlation function
$f(j,k)$ has to fulfill the condition,
\begin{equation}
  \label{eq:fjkCondEq}
    \pe(k) = \sum_j P(j,k) = \pe(k) \, \mean{f(j,k)}_j ,
\end{equation}
which means
\begin{equation}
  \label{eq:fjkCond}
    \mean{f(j,k)}_j = 1  .
\end{equation}
The considerations so far are general. However, as we want to control
correlations within the network, we seek for an explicit correlation
function $f(j,k)$ which has the property of Eq.  \eqref{eq:fjkCond}
and produces a joint degree distribution which yields a given average
nearest neighbor degree $\knn(k)$ function. To do so, we make a simple
ansatz for the correlation function
\begin{equation}
  \label{eq:fjkAnsatz}
  f(j,k) = 1 + h(j) \, h(k) .
\end{equation}
This functional form may be understood as a series expansion of
  first order, fulfilling the necessary symmetry property that the
  correlation function has to be constant under exchange of indices $j$
  and $k$.
Plugging this ansatz into Eq. \eqref{eq:knnFjkRel} takes us to
\begin{equation}
  \label{eq:hkCalcEq}
    \knn(k) = \mean{k} + \mean{j \, h(j)} \, h(k) ,
\end{equation}
which means that
\begin{equation}
  \label{eq:hkCalc}
    h(k) = \frac{\knn(k) - \mean{k}}{\mean{j \, h(j)}} .
\end{equation}
The constant $\mean{j \, h(j)}$ can easily be calculated by
multiplying Eq. \eqref{eq:hkCalc} with $k\, \pe(k)$ and summing over
all $k$. Rearranging the terms then yields
\begin{equation}
  \label{eq:jhjConst}
  \mean{k \, h(k)} = \sqrt{\mean{k \, \knn(k)} - \mean{k}^2 }
  = \sqrt{r \sigma^2_\mr{e}} .
\end{equation}
Finally, the correlation function $f(j,k)$ has the form
\begin{equation}
  \label{eq:fjkKnnFinal}
  f(j,k) = 1 + \frac{1}{r} \frac{(\knn(j) - \mean{k}) \, (\knn(k) -
    \mean{k})}{\sigma^2_\mr{e}} .
\end{equation}
Employing condition
\eqref{eq:fjkCond} to the ansatz in 
Eq.~(\ref{eq:fjkAnsatz}) yields
\begin{equation}
  \label{eq:hjKnnCondBase}
    \mean{h(j)} = 0 .
\end{equation}
This property is consistent with the functional form of $h(k)$ in
Eq.~(\ref{eq:hkCalc}), since the average of $h(k)$ over $k$ with
respect to the edge end distribution $\pe(k)$ yields zero by usage of
Eq.~(\ref{eq:knnMeanProp}) ($\mean{\knn(k)} = \mean{k}$).
Eq.~(\ref{eq:knnMeanProp}) helps furthermore to construct valid
average nearest neighbor functions $\knn(k)$ with an arbitrary
functional dependence upon the degree $k$.  Taking a sufficiently
smooth and positive weighting function $g(k)$, the corresponding
$\knn(k)$ compatible with Eq.~\eqref{eq:knnMeanProp} is then
\begin{equation}
  \label{eq:knnConstr}
  \knn(k) = \frac{\mean{k}}{\mean{g(k)}} \, g(k) .
\end{equation}
However, the resulting correlation function $f(j,k)$ is still
constrained by even further conditions
\cite{J.-S.2006,S.N.2005,M.2004}. For example, the ratio $r_{j,k}$ as
introduced in Ref.~\cite{M.2004} is defined as the actual number of
connections $E_{j,k}$ ($= P(j,k) \bar{k} N$) divided by the maximal
number of connections $m_{j,k}$ among the degree classes $j$ and $k$.
For networks without multiple edges this ratio is given by
\begin{equation}
  \label{eq:rjkDef}
  r_{j,k} = \frac{E_{j,k}}{m_{j,k}} = \frac{P(j,k)}{\min\{ \pe(j) ,
    \pe(k), \bar{k} N \, \pe(j) \pe(k) / j k \}} .
\end{equation}
It is clear that this ratio must always be in the range between $0$
and $1$ for all valid degree classes $j$ and $k$ present in the
network,
\begin{equation}
  \label{eq:rjkCond}
  0 \leq r_{j,k} \leq 1 \; \forall \; j, k \in [\km, \kM] .
\end{equation}
From this condition the admissible degree range $[\km, \kM]$ becomes
dependent upon the details of the correlation function $f(j,k)$. To
proceed, we choose as an example the average nearest neighbor function
to be a power law $\knn(k) \propto k^\alpha$, as this functional form
roughly approximates the measured average nearest neighbor function of
various empirical networks. Using this ansatz, one obtains the final
form of the correlation function as
\begin{equation}
  \label{eq:fjkFinal}
  f(j,k) = 1 + \frac{1}{\mean{k^{\alpha+1}}/\mean{k} -
    \mean{k^\alpha}} \, \frac{1}{\mean{k^\alpha}} \, (j^\alpha
  - \mean{k^\alpha}) \, (k^\alpha - \mean{k^\alpha}) .
\end{equation}
Up to this point the degree distribution $P(k)$ or equivalently the
edge end distribution $\pe(k)$ is still arbitrary as the former does
only enter Eq.~(\ref{eq:fjkFinal}) via the averages $\mean{ \cdot }$
used in the definition of the correlation function $f(j,k)$.
Nevertheless, the range of the exponent $\alpha$ is limited, since
condition of Eq.~(\ref{eq:rjkCond}) has to be fulfilled. A
further complication arises from intrinsic correlations
caused by the constraint of the absence from self- and multiple-edges.
In the following we discuss these issues for scale-free networks and
empirical networks in detail.

\subsection{Scale-Free Networks}
\label{sec:sfNets}

The degree distribution $P(k)$ of a scale-free network is defined by
\begin{equation}
  \label{eq:pkDef}
  P(k) \propto k^{-\gamma},
\end{equation}
where $\gamma$ is the scale-parameter. The edge end distribution is
therefore given by
\begin{equation}
  \label{eq:pkeDefSF}
  \pe(k) \propto k^{-\gamma + 1} .
\end{equation}
As we only discuss finite networks, the range of admissible degrees
$k$ is limited by various conditions. First, the rapidly decreasing
probability for increasing degrees $k$ requires to cut-off the degree
range at a maximal degree $\kM$ above which the accumulated
probability weight is equal to $1/N$. This yields the so-called
natural cut-off \cite{Cohen2000},
\begin{equation}
  \label{eq:4}
  \kM^{\mt{natural}} = N^{1/(\gamma-1)} .
\end{equation}
This natural cut-off is necessary to prevent large fluctuations in a
finite random network ensemble and is an upper limit for the maximal
degree $\kM$. It is important to emphasize that this cut-off is by no
means induced by the topology of the complex network.

However, it turns out that the natural cut-off is not always
compatible with the condition of Eq.~(\ref{eq:rjkCond}), which can
easily be used to determine the so-called structural cut-off. In the
case of scale-free networks, Eq.~(\ref{eq:rjkDef}) reduces for
sufficiently large degrees $j$ and $k$ to $r_{j,k} = j k \, f(j,k) /
\bar{k} N$ and defines therefore a maximal degree $\kM$ at the upper
bound for the ratio ($r_{\kM,\kM} = 1$). With this criteria, one
obtains, in the case of uncorrelated networks having a constant
correlation function $f(j,k) = 1$, the scale-parameter independent
cut-off $\kM^{\mt{structural}} \propto N^{1/2}$. This is smaller than
the natural cut-off for values of the scale-parameter in the range $2
< \gamma \leq 3$ . Nevertheless, newer calculations by
\citeauthor{S.N.2005} \cite{S.N.2005} reveal that this structural
cut-off is still too large in that particular range of the
scale-parameter $\gamma$ and causes intrinsic correlations to arise
within otherwise uncorrelated networks without self- or
multiple-edges.  Due to the maximal degree $\kM$ being too large and
the required constraints, the vertices with large degrees $k$ do have
a tendency to connect preferably with low degree vertices which
effectively yields disassortativity. The reason for the failure of
condition (\ref{eq:rjkCond}) in the case of scale-free networks with a
scale-parameter $\gamma$ in the range $(2,3]$ can be seen in the
diverging fluctuations in the degree distribution as only the first
moment of the degree distribution $P(k)$ is finite. The approach taken
by \citeauthor{S.N.2005} is based upon a statistical ensemble ansatz.
A canonical network ensemble is defined as the set of networks with a
fixed set of vertices and a fixed number of edges. The final networks
are then the out-come of an evolution process where randomly chosen
edges are removed and simultaneously added to a pair of vertices in
the network. The pair of vertices is chosen at random with weights
given by the product of a preferential function $f(j) \, f(k)$ where
$j$ and $k$ are the degrees of the respective vertices. With the
preferential function $f(k) = k + 1 - \gamma$ and beneath the critical
temperature, the authors observe that the degree distribution becomes
scale-free.  However, depending upon the finiteness of the second
moment of the degree distribution, \citeauthor{S.N.2005} find
different cut-offs of the degree range
\begin{equation}
  \label{eq:5}
  \kM^{\mt{ensemble}} =
  \begin{cases}
    N^{1/2} & \mt{if} \quad \gamma > 3 \\
    N^{1/(5-\gamma)} & \mt{if} \quad 2 < \gamma \leq 3 .
  \end{cases} 
\end{equation}
The evolution process driving a network into this equilibrium network
is, of course, neither the same as constructing a network with the CM
algorithm nor with the algorithm developed in this paper. The CM
algorithm and the algorithm presented in this paper, however, fix {\em
  a priori} the number of vertices and edges as well, just as in the
canonical network ensemble. Thus, both algorithms can be interpreted
to produce graphs which are members of the canonical network ensemble
below the critical temperature, since both approaches evidently yield
random networks with the correct degree distribution.

Up to this point, we have only treated the uncorrelated case which
corresponds to $\alpha = 0$ in Eq.~(\ref{eq:fjkFinal}). Numerical
experiments indicated a strong deviation from the expected power-law
for the measured average nearest neighbor $\knn(k)$ function in the
case of assortative networks which have $\alpha > 0$, if one naively
uses a cut-off as it is applicable for uncorrelated networks. The
average nearest neighbor function shows that the vertices with the
largest degree fall below their expected average nearest neighbor
value and tend therefore to cause some degree of disassortivity. This
effect roots in the constraint of the prevention of self- and
multiple-edges and becomes stronger for larger values of the exponent
$\alpha$. To compensate this effect, we incorporated the exponent
$\alpha$ in the exponents of the maximal degrees identified so far in
a simple way (an analytically exact derivation is beyond the scope of
this paper) and always use the minimal resulting maximal degree,
\begin{equation}
  \label{eq:kMaxFinalDef}
  \kM = \min\{ N^{(1-\alpha)/(5-\gamma) \,  },
  N^{(1-\alpha)/(\gamma-1)}, N^{1/(\gamma-1)}\} .
\end{equation}
Using a maximal degree of this form lowers (raises) the cut-off degree
for assortative (disassortative) correlations with increasing
(decreasing) exponent $\alpha$. Having fixed the maximal degree $\kM$,
we set the minimal degree $\km$ to be $2$ in all simulations. This
ensures that we always obtain a largest giant component in the network
having almost the size of the entire network, which in turn guarantees
that the largest giant component has the same two-point correlation
structure as the entire network. This is favorable, since in most
applications only the largest component of the generated random
networks is of interest.

\begin{figure}[t]
  \centering
  \includegraphics{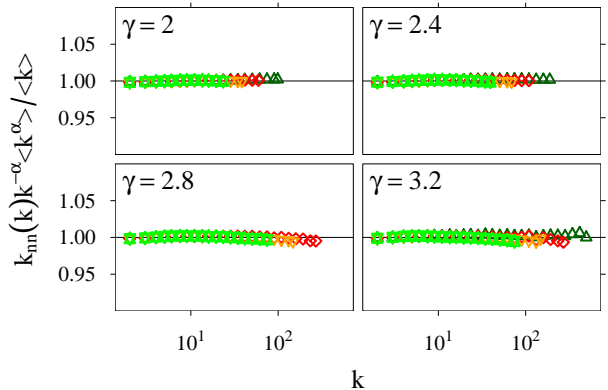}
  \caption{(color online) Data-collapse for average nearest neighbor
    function $\knn(k) \propto k^\alpha$ with various values of the
    power parameter $\alpha$ for networks with a scale-free degree
    distribution with varying values of the scale parameter $\gamma$.
    The symbols used for the different values for the scale-parameter
    $\alpha$ are: blue circle $-0.2$, pink square $-0.1$, dark green
    triangle up $0.0$, red diamond $0.1$, yellow triangle down $0.2$,
    and light green star $0.3$.}
  \label{fig:knnTest}
\end{figure}

\begin{figure}[t]
  \centering  
  \includegraphics{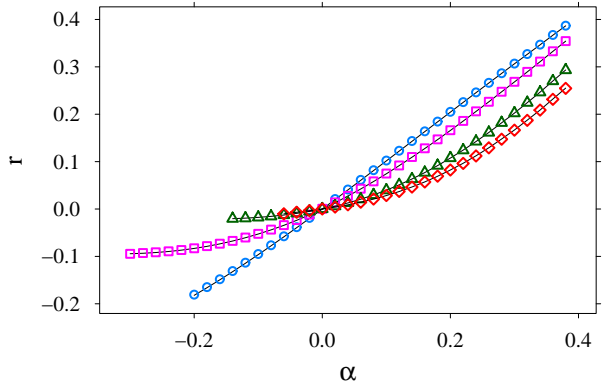}
  \caption{(color online) Newman factor $r$ as a function of the
    scale-parameter $\alpha$ for different values of the
    scale-parameter $\gamma$. The straight line denotes the theoretic
    values of the Newman factor $r$ as of Eq.~(\ref{eq:knnNewmanRel}).
    The symbols denote the value of the scale-parameter $\gamma$: blue
    circle $2.0$, pink square $2.4$, dark green triangle $2.8$, and
    red diamond $3.2$.}
  \label{fig:knnTestAlpha}
\end{figure}

As already pointed out, it is crucial to note that only the first
moment of the degree distribution is finite for values of the
scale-parameter $\gamma$ in the range $(2,3]$ while all higher moments
diverge. However, already the first moment of the edge end
distribution $\pe(k)$ is diverging in this range of the
scale-parameter $\gamma$.  This has the important consequence that the
average nearest neighbor function $\knn(k)$ becomes system size
dependent, as $\mean{\knn(k)} = \mean{k}$ by
Eq.~(\ref{eq:knnMeanProp}).  To validate the predicted power-law
behavior of the average nearest neighbor function $\knn(k)$, we employ
a dimensionless data-collapse of the function,
\begin{equation}
  \label{eq:knnCollapse}
  \knn(k) \, k^{-\alpha} \, \mean{k^\alpha}/\mean{k} = 1 .
\end{equation}
This type of plot is extremely sensitive even against smallest
deviations from the predicted power-law in the average nearest
neighbor function $\knn(k)$. The numerical results for various values
of the scale-parameters $\gamma$ and the exponent $\alpha$ are shown
in Fig.~\ref{fig:knnTest} for networks of size $N = 10^6$. Each data
point is calculated over an ensemble of $10^3$ random networks.  The
curves run quite nicely along the predicted constant line of $1$.
Especially the $\alpha = 0$ curves coincide with the constant line of
$1$, which is a further, very important validation of the algorithm,
since in this case the algorithm has to coincide with the well-known
UCM algorithm \cite{Catanzaro2005}. Three details are interesting to
note: (i) with decreasing $\alpha$ the curves become longer as the
maximal degree $\kM$ increases, (ii) not all values of the exponent
$\alpha$ can be realized for a given value of the scale-parameter
$\gamma$ as condition $r_{j,k} \geq 0$ is violated for some curves and
would require a further adjustment of $\kM$ or even $\km$, (iii) with
increasing scale-parameter $\gamma$ the curves for larger values of
the exponent $\alpha$ show a trend to slightly bend below the constant
line of $1$ which is an indication that the cut-off as of
Eq.~(\ref{eq:kMaxFinalDef}) still gives slightly too large values for
the maximal degree $\kM$.  Another test of our formalism can be
accomplished by comparing the Newman factor $r$ of the resulting
networks to the values of the analytically predicted ones by
Eq.~(\ref{eq:newmanFjkRel}). The Fig.~\ref{fig:knnTestAlpha} shows
that numerical simulations (points) and theoretical predictions
(lines) coincide very well.

\begin{figure}[t]
  \centering
  \includegraphics{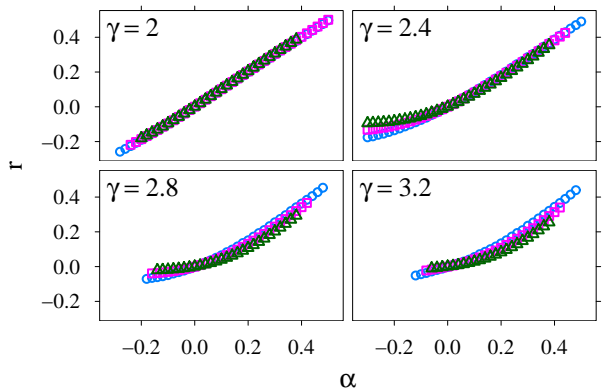}
  \caption{(color online) Network size dependence of the Newman
    factor $r$ as a function of the exponent $\alpha$ for
    different values of the scale-parameter $\gamma$. The network size
    $N$ is marked by the symbols: blue circle $10^4$, pink square $10^5$, and
    dark green triangle $10^6$.}
  \label{fig:knnTestAlpha_Nvar}
\end{figure}

The diverging moments $\mean{k}$ and $\mean{k^{\alpha+1}}$ of the edge
end distribution $\pe(k)$ for values of the scale-parameter $\gamma$
within the range $(2,3]$ make a careful inspection of finite-size
effects necessary. One can easily see that the ratio
$\mean{k^{\alpha+1}}/\mean{k}$, appearing in the denominator of the
correlation function $f(j,k)$ in Eq.~(\ref{eq:fjkFinal}), diverges, as
the ratio becomes proportional to $\kM^\alpha$. Nevertheless, a
detailed calculation reveals certain restrictions on the maximal range
of admissible degrees $k$ if $\alpha$ is chosen to be different than
$0$. In this case, the criterion $r_{j,k} \geq 0$ leads to a relation
between the minimal degree $\km$ and the maximal degree $\kM$. Thus,
the range of admissible degrees is limited and the moments $\mean{k}$
and $\mean{k^{\alpha+1}}$, which would otherwise diverge, remain
finite. Fig.~\ref{fig:knnTestAlpha_Nvar} shows the finite-size
effects on the Newman factor $r$ as a function of the exponent
$\alpha$.  The plot shows only a marginal effect of the system size
$N$ on the curves. However, for smaller sizes, a broader range in the
exponent $\alpha$ can be used. This is due to a violation of the
$r_{j,k} \geq 0$ criterion which requires for larger networks either a
smaller maximal degree $\kM$ than the one used from
Eq.~(\ref{eq:kMaxFinalDef}) or a greater minimal degree $\km$. Despite
the restrictions which apply to the ansatz made, the range of
correlations span very well the range of correlations found in
empirical networks.

\subsection{Empirical Networks}
\label{sec:empNets}

\begin{figure}[t]
  \centering
  \includegraphics{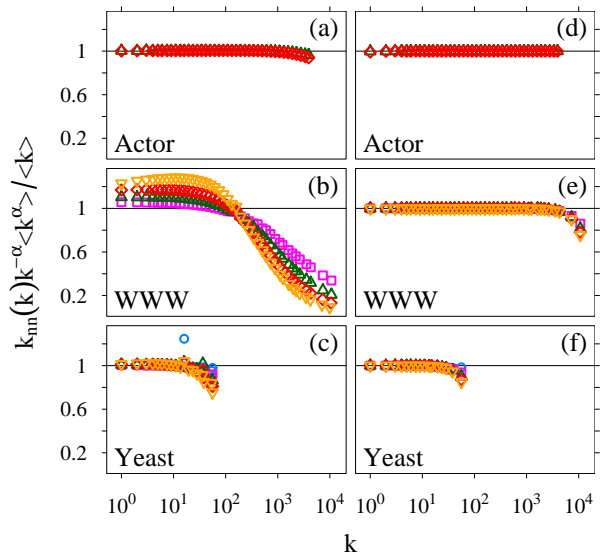}
  \caption{(color online) Data-collapse for average nearest neighbor
    function $\knn(k)$ for the three empirical networks
    actor-, WWW- and yeast protein-interaction network. The left
    column (a), (b), and (c) shows the data-collapse for networks
    generated by the algorithm, while the right column (d), (e), and
    (f) shows the same data-collapse for networks simulated in an
    annealed manner. The statistics for each curve is $10^2$, $10^3$,
    and $10^4$ realizations, respectively. The different symbols
    indicate different values for the exponent $\alpha$: blue circle
    $-0.2$, pink square $-0.1$, dark green triangle up $0.0$, red
    diamond $0.1$, and yellow triangle down $0.2$.}
  \label{fig:knnTestEmpirical}
\end{figure}

A very interesting aspect of our formalism is its applicability to
empirical networks. By extracting a degree sequence from an empirical
network and employing the formalism developed in the last section, it
is possible to create random networks which have the same degree
sequence as the empirical network and an arbitrarily chosen average
nearest neighbor function $\knn(k)$, for instance following a
power-law with tunable exponent $\alpha$. Thus, given a degree
sequence from a network, one constructs from this the corresponding
edge end distribution $\pe(k)$ and calculates then via
Eq.~(\ref{eq:fjkFinal}) a joint degree distribution $P(j,k)$ with
which one builds a randomized network. As a result, one obtains
randomized versions of the empirical network with freely tunable
two-point correlation strength, depending upon the choice of the
exponent $\alpha$. However, the range of the exponent $\alpha$ is
limited by condition~(\ref{eq:rjkCond}). In
Fig.~\ref{fig:knnTestEmpirical} (a), (b), and (c) the numerical
results are shown for the actor-, the WWW-, and the yeast-network. The
plot uses the same type of data-collapse as already presented in
Fig.~\ref{fig:knnTest}.  The deviations from the expected constant
value of $1$ for the data-collapse are due to intrinsic correlations
which arise in networks without neither self- nor multiple-edges and
are caused by the maximal degree $\kM$ in the degree sequence (see
section \ref{sec:sfNets}). Especially the WWW-network is strongly
affected by this as it has a maximal degree $\kM$ of the order $10^4$,
while the network size is $10^5$ and hence only one order of magnitude
greater.

\section{Annealed Networks}
\label{sec:annealed}

To investigate, for example, a dynamical processes on random networks,
one typically performs the dynamical process on a whole ensemble of
networks and computes averages of the observables one is interested
in. The algorithm presented so far is suitable to generate such random
network ensemble. The network itself always stays constant during one
dynamical process and one refers to this type of network typically as
static or quenched network. A different approach is to change the
network on a certain time-scale during a dynamical process and then
calculate averages over time of the observables one is interested in.
In an extreme case, the vertices of the network are reshuffled before
every microscopic step of the dynamic. Such changing networks are
referred to as annealed networks (see
Ref.~\cite{Burda2001,dorogotvtsev-evolnets03,Dorogovtsev2003,Stauffer2005}).
If the dynamic is local in each microscopic step (for instance a
diffusion step from one vertex to another along an edge), it is
sufficient to draw edges on demand only and to generate solely the
local connections around the vertex considered. Here, we propose a
scheme which efficiently simulates such annealed networks . The idea
is to treat vertices of a network discrete while the edges are solely
represented by an arbitrary joint degree distribution $P(j,k)$ such
that the connectivity structure of the network is only defined on
average. Hence, this scheme effectively simulates the networks
connectivity structure in a mean field (MF) like manner.

This is a very convenient tool as theoretical approaches to complex
network topics are frequently based on MF theories. Successful
examples are reaction-diffusion systems
\cite{Catanzaro2005a,Weber2006}, epidemic disease spreading
\cite{Bogun'a2003}, and phase transitions in ferromagnetic magnets
\cite{Dorogovtsev2002a}, to mention just a few examples. These
theories usually describe the network topology via a statistical
approach. Thus, it is desirable to numerically represent networks in a
probabilistic manner as well.  This allows an even better test of MF
based theories since the network is represented as it is done within
the theory. Furthermore, by comparison of quenched with annealed
simulations, one can analyze in detail which aspects of such a MF
theory are an over-approximation due to the MF assumption. We define
such an annealed network to consist of a degree sequence
$\{k_i\}$ of size $N$ and a corresponding joint degree distribution
$P(j,k)$. Each element $i$ of the degree sequence represents a vertex
with $k_i$ connections. Thus, the set of edges is not fixed, only the
total number of edges ($N_\mt{e} = \sum_i k_i$) is held constant.
Whenever, for example, a dynamical process requests an adjacent vertex
of a given vertex, the neighbor vertex is instantly determined by
sampling one edge which emanates from the given vertex. This edge is
drawn from the joint degree distribution $P(j,k)$ and will instantly
be removed after usage.

This simulates a continuously rewired network which is only locally
defined by means of one edge at a time.  The first four steps to setup
such an annealed network are basically the same as done for the
initialization of the algorithm of section~\ref{sec:algorithm}: (i)
Draw a degree sequence from the joint degree distribution $P(j,k)$ or
take the degree sequence from a real network. That degree sequence is
(ii) sorted according to degree classes and (iii) mapped into a
discrete edge end distribution $\ped(k)$. In the same manner as done
previously, (iv) one calculates the discrete conditional degree
distribution $\pjkcd$ from the theoretical joint degree distribution
$P(j,k)$. Now, instead of constructing the network, one only redefines
how neighbors of vertices and hence how edges have to be understood:
\begin{itemize}
\item The neighbor vertices of a vertex with degree $k$ are always
  drawn by the conditional probability distribution $\pjkcd$.
\item An edge is sampled by first drawing a vertex via the edge end
  distribution $\ped(k)$ and secondly the vertex neighbor is found by
  sampling the conditional probability distribution $\pjkcd$.
\end{itemize}
As we want the network to be free of self-connections,
we assure that the sampled vertices at both ends of the sampled edges
are not the same. However, the constraint of preventing multiple-edges
among vertices is not possible to be enforced within this local
definition of the network. Therefore, these annealed networks are free
of the intrinsic degree correlations which arise due to this
particular constraint. This becomes apparent in
Fig.~\ref{fig:knnTestEmpirical}(d), (e), and (f) where numerical
results of annealed networks are shown as a data-collapse for the
average nearest neighbor function $\knn(k)$, aside with the
corresponding curves in the case where the network is actually
constructed (Fig.~\ref{fig:knnTestEmpirical}(a), (b), and (c)). Only
the curve for the WWW network, Fig.~\ref{fig:knnTestEmpirical}(e),
deviates from the expected value of $1$ for very large degrees. This
has to be attributed to the prevention of self-connections, which is
still enforced.  Since these vertices with a very large degree are not
allowed to connect to themselves, they have to connect on average with
vertices which have a degree below the preassigned average nearest
neighbor function $\knn(k)$, causing some slight trend towards
disassortativity.

\section{Conclusions}
\label{sec:conclusion}

In summary, we have presented an efficient and accurate algorithm
which generates networks with an {\em a priori} defined two-point
correlation structure defined by an arbitrary joint degree
distribution $P(j,k)$. This provides much better null models for the
investigation of empirical networks, as these are usually two-point
correlated.  Besides the applicability to reconstruct the two-point
correlations of empirical networks, we developed a formalism which
allows to systematically tune the strength of two-point correlations
in a network while preserving the degree distribution $P(k)$ of a
network. The two-point correlations are specified in our ansatz via
the average nearest neighbor function $\knn(k)$ which we exemplified
by a power-law ansatz $\knn(k) \propto k^\alpha$ with the tunable
exponent $\alpha$. As two important examples, we employed this
formalism in the cases of scale-free networks and empirical networks.
However, as intrinsic degree correlations arise from the constraint of
the prevention of self- and multiple-edges, these cause inevitable
deviations from the theoretically preassigned two-point correlations.
Furthermore, we found that the maximal cut-off degree $\kM$ in the
case of articifial scale-free networks to prevent these intrinsic
correlations is substantially lower than it was believed.

At last, we introduced the notion of two-point correlated annealed
networks which are ideally suited to test the validity of mean field
theories, since the edges of these networks are solely represented in
a probabilistic manner.

Using this algorithm and the new formalism developed, one can
investigate the effects of two-point correlations in empirical and
artificial networks. Such scheme is expected to be an important tool
to better understand, for example, how the topology of a network
influences dynamical processes on it.

\bibliography{gen_corNets}

\end{document}